\pdfoutput=1

\documentclass[11pt]{article}

\usepackage[]{ACL2023}
\usepackage{makecell}
\usepackage{times}
\usepackage{latexsym}
\usepackage{graphicx}
\usepackage{multirow}
\usepackage{makecell}
\usepackage[T1]{fontenc}

\usepackage[utf8]{inputenc}

\usepackage{microtype}

\usepackage{inconsolata}

\title{Reviewriter: AI-Generated Instructions For Peer Review Writing}
\author{Xiaotian Su\textsuperscript{1}, Thiemo Wambsganss\textsuperscript{1}, Roman Rietsche \textsuperscript{2},\\ \textbf{Seyed Parsa Neshaei\textsuperscript{1}}, \textbf{Tanja Käser\textsuperscript{1}}\\ \textsuperscript{1} EPFL, Lausanne, Switzerland\\
\footnotesize{ {\tt \{xiaotian.su, thiemo.wambsganss, seyed.neshaei, tanja.kaeser\}{\tt@epfl.ch}}}\\
 \textsuperscript{2} Universtiy of St.Gallen, St.Gallen, Switzerland\\
  \footnotesize{ {\tt {roman.rietsche@hsg.ch}}}\\
\\}

\graphicspath{{figures/}}
\begin{document}
\maketitle
\begin{abstract}
Large Language Models (LLMs) offer novel opportunities for educational applications that have the potential to transform traditional learning for students. Despite AI-enhanced applications having the potential to provide personalized learning experiences, more studies are needed on the design of generative AI systems and evidence for using them in real educational settings. In this paper, we design, implement and evaluate \texttt{Reviewriter}, a novel tool to provide students with AI-generated instructions for writing peer reviews in German. Our study identifies three key aspects: a) we provide insights into student needs when writing peer reviews with generative models which we then use to develop a novel system to provide adaptive instructions b) we fine-tune three German language models on a selected corpus of 11,925 student-written peer review texts in German and choose German-GPT2 based on quantitative measures and human evaluation, and c) we evaluate our tool with fourteen students, revealing positive technology acceptance based on quantitative measures. Additionally, the qualitative feedback presents the benefits and limitations of generative AI in peer review writing.
\end{abstract}

\section{Introduction}

\begin{figure*}[!h]
    \centering
    \includegraphics[width=\linewidth]{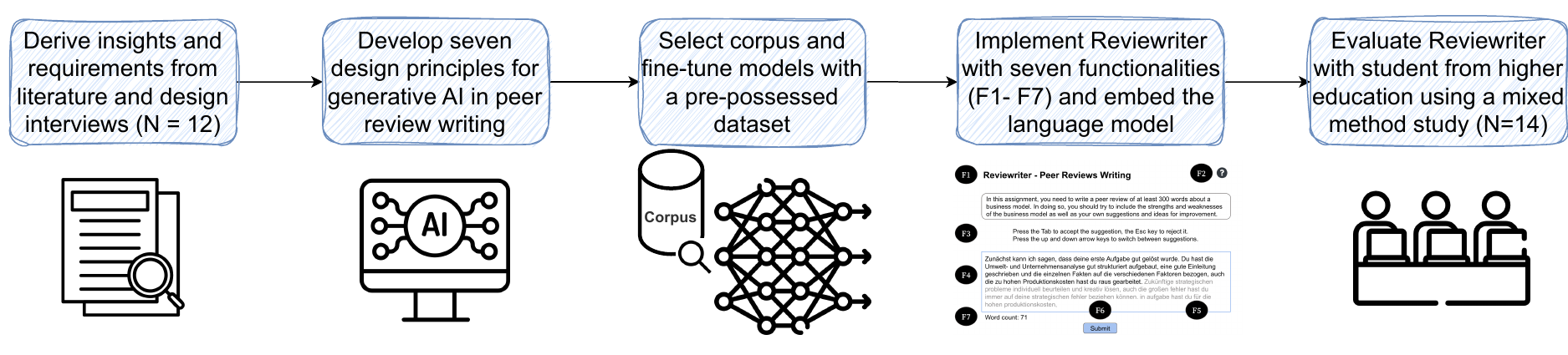}{
        \caption{Overview of our methodology: We first gather system needs and requirements from literature and student interviews. Then we derive seven design principles with pedagogical considerations for a tool to provide AI-generated instructions for peer review writing tasks. Next, we fine-tuned three language models based on a selected corpus \cite{wambsganss-bias-review-corpus}. Then, we instantiate the design in \texttt{Reviewriter} and evaluate it with fourteen students to assess its performance and gather quantitative as well as qualitative feedback.}
        \label{fig:methodology}
    }
\end{figure*}
Peer reviewing is a process by which learners provide formative feedback to each other on an individual task based on assessment criteria \cite{sadler2006impact, Rietsche.2019b}. Research has found theoretical and empirical evidence for the positive effects of peer reviews on critical thinking skills \cite{lin2021facilitating, ibarra2020developing}, communication skills \cite{lai2016training}, and learning motivations \cite{16motivation}. The prevailing practice of peer review in tertiary education is evident in the eruption of massive open online courses (MOOCs) \cite{li2016peer}. In these large-scale learning scenarios, peer review is particularly important since it is challenging for teachers to give effective one-by-one feedback due to immersive workload and shortage of time \cite{er2021collaborative}. However, according to \citet{oliver1982helping}, a challenge that plagues many student writers, including those having satisfactory grammar and spelling skills, is writer's block. It was defined by \citet{1980rigidblock} as "that frustrating, self-defeating inability to generate the next line, the right phrase, the sentence that will release the flow of words again." A collaborator who provides instructions and points out new directions might help alleviate writer’s block \cite{clark2018creative} and the combination of a writer’s own ideas with suggested ideas is a form of psychological creativity \cite{boden2004creative}. Novel LLMs have the potential to address the challenge of writer's block by generating suggestions for the next lines, right phrases, or sentences, thereby facilitating the flow of ideas \cite{2022sparks}, and helping students compose responses more efficiently \cite{van2023chatgpt, 21haichatbot}. There are LLM-based collaborative writing tools to provide support for various writing tasks, including story writing \cite{yang2022ai}, science writing \cite{2022sparks}, and screenwriting \cite{mirowski2022co}. However, few have investigated the utilization of generative AI for peer review writing tasks. Therefore, in this paper, we build and evaluate \texttt{Reviewriter} which can provide AI-generated instructions tailored to students' needs while writing peer reviews. It suggests possible directions based on students' input to inspire divergent outcomes while still leaving learners in control of the final text. 

To investigate how to provide students with help to overcome writer's block in peer review writing, we conduct a literature review to gather insights for a peer review support system. We summarize five user requirements from interviews with twelve graduate students. Based on those, we develop seven design principles for providing AI-generated instructions in peer review tasks. Next, we search peer review corpora satisfying certain criteria and pre-process 11,925 student-written peer review texts in German \cite{wambsganss-bias-review-corpus}. We use it to fine-tune three language models to provide students with informative instructions. The best results according to training loss and human evaluation of fluency and correctness are achieved by German GPT-2. Then, we implement the design principles into the system to provide AI-generated instructions for peer review writing. Finally, in a mixed-method study with our full-working prototype, we evaluate the performance of the tool in a real-world learning exercise with fourteen students, and four of them also participated in the design interview. We assess the technology acceptance and level of enjoyment of the tool using well-defined constructs from \citet{venkatesh2008technology, venkatesh2003user} and also collect qualitative feedback from students. 

Our research makes three contributions to the innovative use of NLP in education. Firstly, we provide insights and practical design considerations for incorporating AI-generated instructions in peer review writing tasks to overcome the known challenge of writer's block \cite{oliver1982helping}. Secondly, we present and compare three open-source language models fine-tuned on a selected corpus of 11,925 student-written peer review texts in German. Lastly, we build \texttt{Reviewriter}, which implements seven functionalities with pedagogical design considerations and evaluates it on fourteen students from tertiary education. Our findings suggest that the tool providing AI-generated instructions in students' peer writing tasks leads to high ease of use and a high intention to use for students in their review writing process. Moreover, in the qualitative feedback, we find that the model has the potential to provide novel ideas for students to continue in depth. However, like other LLMs, it suffers from hallucination \cite{maynez20faithfulness} by producing factually incorrect and nonsensical answers, this invites further research to overcome and mitigate
artificial hallucination. With \texttt{Reviewriter}, we present an interface with design rationales and an evaluated tool that other researchers can build upon to explore the effects of LLMs and the benefits and limitations of generative AI for writing peer reviews and building educational applications.

\section{Related work}

\subsection{Student peer reviewing}
There has always been significant interest in the study of peer reviews in the NLP community. \citet{jia22starting} introduced an approach called incremental zero-shot learning (IZSL) to address the issue of insufficient historical data for peer reviews. \citet{wambsganss22alen} used empathy detection algorithms from NLP to analyze the given text and provide adaptive feedback in students’ peer writing process. Moreover, several works have investigated how to embed classification models to support students in peer review writing. For example, researchers have explored the use of these models to develop argumentation skills \cite{wam20al}, support cognitive and emotional empathy writing \cite{wambsganss2021supporting}, and assess the specificity of written peer feedback \cite{rietsche22specificity}. While NLP models, particularly LLMs, have the potential to deliver adaptive learning content \cite{adiguzel2023revolutionizing, qadir2022engineering}, little research has focused on how to leverage their ability to provide tailored instructions for students during peer review writing \cite{darvishi2022}. \citet{van2023chatgpt} mentioned benefits provided by generative AI for completing peer review tasks quickly. Experimental results from \citet{21haichatbot} showed that the effectiveness of generated suggestions, regardless of their performance quality, has consistently helped humans compose responses more efficiently when providing suggestions. In addition, \citet{2022sparks} demonstrated that students find it faster and easier to draw on language from generated texts than to write a sentence from scratch, even when given well-known information. Therefore, we propose a novel peer review writing tool \texttt{Reviewriter}, by leveraging the power of generative models, it can provide students with adaptive instructions to help them overcome writer's block in peer review writing.

\subsection{NLP for writing support}
With the massive success of ChatGPT, NLP is rapidly evolving as a key tool in writing support. On one hand, there is widespread adoption of generative AI in practice. Commercial writing assistants like Monica \footnote{https://monica.im/}, a ChatGPT-powered extension, can support copywriting. And specialized applications like Jenni AI \footnote{https://jenni.ai/}, Jasper AI \footnote{https://www.jasper.ai} and Notion AI \footnote{https://www.notion.so/product/ai} can support creative writing. They are not only able to complete sentences but also generate the whole blog post and many other types of content including essays, emails, stories, and speeches based on users' input. On the other hand, many studies have focused on the use of language models for writing support in tertiary education. For instance, researchers have explored the use of these models for academic writing \cite{2022sparks}, fiction writing \cite{yang2022ai}, and text summarization \cite{dang22textsum}. Despite the widespread adoption of NLP in writing instruction, many models, including ChatGPT, remain general-purpose tools that have not been fine-tuned for specific tasks \cite{chen2023artificial} or designed for particular educational settings \cite{kuhail2023interacting}. Embedding the AI techniques in a student-centered design is a complex task with several socio-technical challenges \cite{xu2021human}, including data collection \cite{Zawacki19ai}, potential bias \cite{adiguzel2023revolutionizing} or discrimination \cite{Pedr2019ArtificialII} in the data, inadequate dataset training \cite{kuhail2023interacting}, incorporating the models, lack of student involvement in the design process \cite{verleger2018pilot}, lacking feedback on the generative system \cite{kuhail2023interacting}, and evaluating student perceptions \cite{xu2021human}. The present work provides insights into how to embed generative AI into peer review writing by establishing student-centered design with pedagogical considerations. We carefully select an unbiased corpus with a sufficient amount of peer review text to fine-tune language models. Furthermore, we evaluate student perceptions quantitatively and collect qualitative feedback on the generative AI system.
\begin{figure*}[!h]
    \centering
    \includegraphics[width=\linewidth]{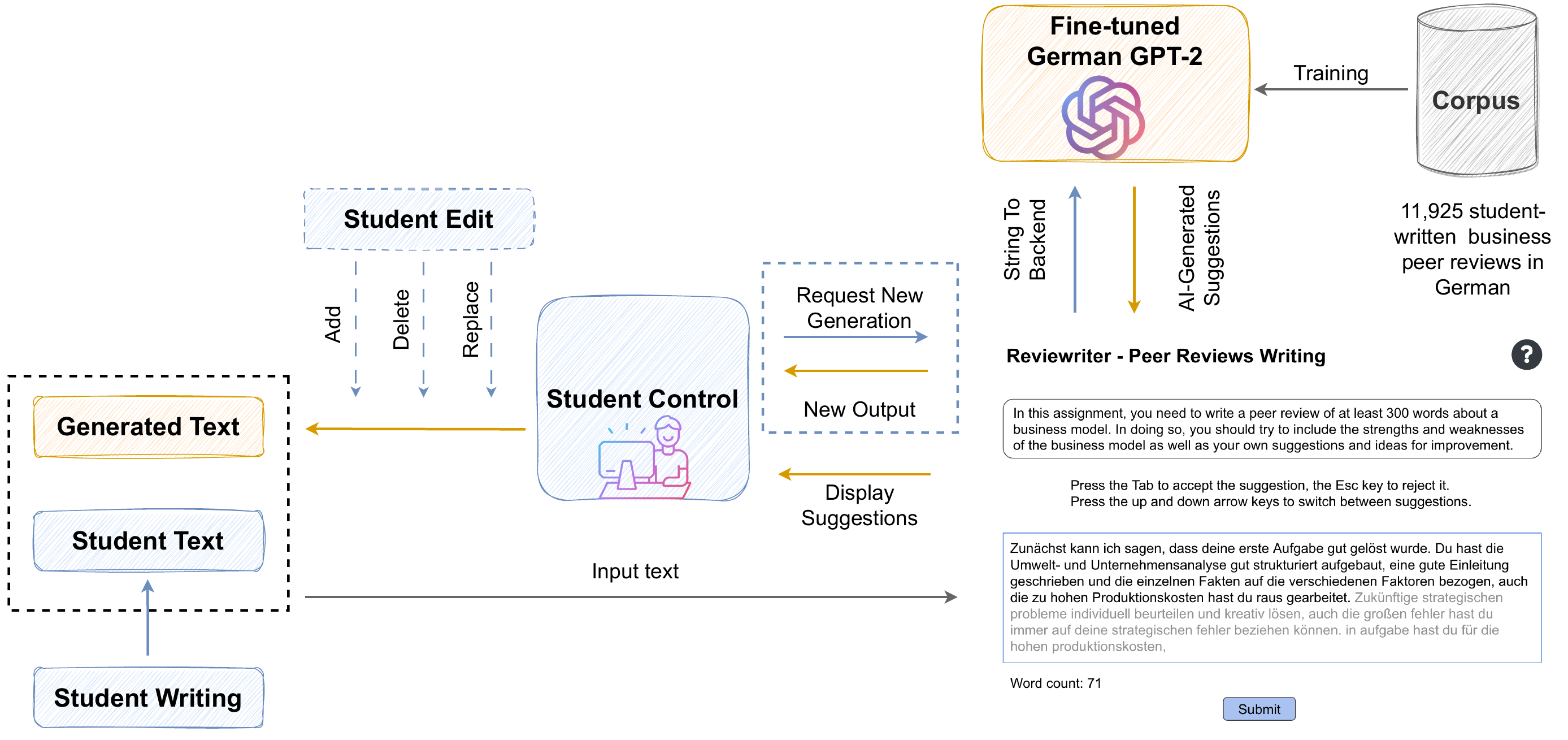}{
        \caption{Architecture of \texttt{Reviewriter} to provide AI-generated instructions for students to write peer reviews. First, students enter initial input, which is then used by the German GPT-2 model to generate  instructions. The students evaluate the generated content and decide whether to regenerate it. Following this, students are free to edit the instructions. Finally, both the generated text and the student's text are utilized as inputs for the next generation.}
        \label{fig:architecture}
    }
\end{figure*}
\section{Generative modeling to provide students adaptive instructions}
\subsection{The peer review dataset}
\label{sec:dataset}
To make sure our system is skilled in providing adaptive instructions for writing peer reviews and to improve accuracy and efficiency for human-AI interaction \cite{lee2022evaluating}, we decide to fine-tune language models with a peer review dataset. We start by searching the literature for a corpus that fulfilled the following criteria: a) it contains a large amount of student-written text in one particular domain (e.g., business model feedback) \cite{kuhail2023interacting}, b) it consists of a sufficient size to represent different nuances of characteristics in a balanced fashion (e.g. specificity, helpfulness) \cite{rietsche22specificity}, and c) it does not possess a significant bias (e.g. gender, racial or social discrimination) \cite{adiguzel2023revolutionizing}. The business model peer review corpus published in \citet{wambsganss-bias-review-corpus} fulfilled all these requirements. The corpus consists of 11,925 peer reviews collected at a university in the German-speaking area of Europe. They were written by first-year master's students in a business department course. The student population has an average age of 24.6 years old with a standard deviation of 1.7 years. Students wrote approximately 9 peer reviews per course with an average length of 220 words. Furthermore, \citet{wambsganss-bias-review-corpus} showed that this collected corpus does not reveal many biases in nine WEAT co-occurrence analyses or in the GloVe embeddings. This corpus provides us with a sufficient amount of unbiased peer review texts to fine-tune language models for adaptive instructions in the domain of business peer reviews.
\subsection{Data pre-processing}
To ensure the model could generate high-quality instructional text, we select reviews written from 2016 to 2021 with a rated helpfulness score greater than five on a 1 - 7 Likert Scale (1: low, 4: neutral, 7: high). We start by removing HTML tags, irrelevant information like PDF file names and specific information like URLs, keywords (revealing the identity of students), and questions asked to write reviews which some students copied to their review text (Appendix \ref{appendix:tq}). We also expand abbreviations as shown in Appendix \ref{appendix:abb}. Then, we shuffle and divide cleaned data into train and test datasets with proportions of 0.8 and 0.2 for fine-tuning and evaluating the language model. Lastly, all sentences are tokenized with model-specific tokenizers.

\subsection{The generative models}
\label{sec:model}
Transformer-based language models, such as BERT \cite{19bert} and GPT-2 \cite{gpt2}, using the pre-training and fine-tuning paradigm, have revolutionized NLP and achieved state-of-the-art records on various tasks. These models are first pre-trained in a self-supervised fashion on a large corpus and fine-tuned for specific downstream tasks \cite{18glue}. In our case, to provide AI-generated instructions for German peer review writing, we use pre-trained causal language models on the HuggingFace platform \cite{wolf-etal-2020-transformers} for German text generation. We choose them because there is no usage limitation and by utilizing open-source technology, we contribute to LLM transparency \cite{van2023chatgpt, adiguzel2023revolutionizing}, allowing other researchers to easily replicate our findings or build upon them. Therefore, we selected two German GPT-2 models (\texttt{dbmdz/german-gpt2} \footnote{https://huggingface.co/dbmdz/german-gpt2} and \texttt{benjamin/gerpt2-large} \footnote{https://huggingface.co/bigscience/bloom-560m}) and one multilingual model BLOOM \cite{2022bloom} (\texttt{bigscience/bloom-560m} \footnote{https://huggingface.co/benjamin/gerpt2-large}). We did not use GPT3 for fine-tuning since it was not open-source available at the time of our research. For all of them, we fine-tune the pre-trained models following the default hyperparameter settings (Appendix \ref{appendix:hyper}) with block size 128, and 500 warm-up steps. 

We compare training loss and used human evaluation to select the best model. Note that GerPT2-large already performs well (Appendix \ref{appendix:text_en} for sample generated text) after ten epochs of training, even with higher training loss compared to the other two models (Table \ref{tab:models}). However, it suffers a long inference time (a student needs to wait around 10 seconds to get instructions given 40 words) compared to the other two models (5 seconds with the same input). Therefore, we decide to further evaluate German GPT-2 and BLOOM. We conduct a human evaluation of the quality of the generated response. Specifically, we sample ten instructions generated by each model and present them to two German researchers to evaluate their fluency and correctness. From the evaluation of both parties, German GPT-2 yields more coherent results than the BLOOM model and there are more meaningless sentences from the response generated by BLOOM than by German GPT-2. Therefore, we decide to use the German GPT-2 model as the base for the tool with a default temperature of 1.0 for generating the next token.

\begin{table}[!h]
\begin{center}
\begin{tabular}{|c|c|c|c|}
\hline
PLM & \makecell{Size \\ \# Param.} & \makecell{Training \\loss} & \makecell{Training \\epochs} \\
\hline
German GPT-2 & 124 & 0.0418 & 30 \\
\hline
BLOOM & 560M & 0.0560 & 30 \\
\hline
GerPT2-large & 774M & 2.8183 & 10 \\
\hline
\end{tabular}
\caption{\label{tab:models}Comparison of the number of parameters for three transformer-based pretrained language models (PLMs) and their training and evaluation loss.}
\end{center}
\end{table}

\subsection{The generative system}
To design a system providing AI-generated instructions for peer review writing, we first draw on insights from relevant literature. Following the methodology of \citet{cooper1988organizing}, we analyze human-AI interaction \cite{shen2023parachute, chan2023mapping, lee2022evaluating} and NLP-supported peer review systems \cite{alqassab2023systematic, darvishi2022}. Then, to gather insights into the needs of writing peer reviews with AI-generated instructions for tertiary education, we conduct semi-structured interviews with twelve graduate students. We reach out to a group of computer science students who previously registered in a business class and have experience writing peer reviews on business models, and to students in our university for general recruitment. The participants have a diverse background in computer science, business, or psychology, and a mean age of 24.50 years (SD = 2.02), including two females and ten males (representing the distribution of computer science students at our school). Half of them had experience writing peer reviews, while the others did not. Each interview lasts around 30 to 50 minutes. We use the expert qualitative interview method outlined in \citet{2013qualitative} and \citet{glaser2009expert-interviews} to gain an initial understanding of students' needs for receiving adaptive instructions in peer review writing. We ask topics about prior experience with technology-based writing systems, perceptions of existing writing systems (e.g., Grammarly), difficulties in writing peer reviews, and desired functionalities for a system to support peer review writing. We transcribe the interviews and identify five clusters of requirements following \citet{cohn2004user}. We find that 75\% of the students would like to interact with a clean and straightforward interface (\textit{user requirement - UR 1}). Two-thirds of interviewees asked for intuitive guidance on how to interact with the tool (\textit{UR 2}). And 41.7\% of them said that they would like to see more than one instruction to choose from (\textit{UR 3}). One-third of the students stated that they prefer to view a complete piece of instruction rather than words or phrases to formulate a concrete idea (\textit{UR 4}). Lastly, two-thirds of them indicated that they would like to see the number of words they have entered to have better control over the structure of the review (\textit{UR 5}). 

\begin{table}[!h]
\begin{center}
\begin{tabular}{ | p{0.1\linewidth} | p{0.8\linewidth} | } 
  \hline
  & Design Principle \\ 
  \hline
  DP1) & Provide a web-based application with a responsive clean and intuitive interface to allow students to use the tool with ease and stay motivated to write. \\ 
  \hline
  DP2) & Provide clear and detailed guidance to ensure that students understand how to use the tool and can take full advantage of the features offered. \\ 
  \hline
  DP3) & Provide an intuitive keyboard control to make it easy for students to manipulate the AI-generated instructions. \\ 
  \hline
  DP4) & Provide a simple text area for students to write, edit the peer review, and view multiple inline instructions. \\
  \hline
  DP5) & Present instructions in an inline format in the text area to help students quickly pick up ideas while allowing them to stay in the context of writing to reduce cognitive burden. \\
  \hline
  DP6) & Provide a complete argument for each instruction to assist students in constructing comprehensive reviews. \\
  \hline
  DP7) & Present a summary of statistics on the text to guide students on how many words they have written. \\ 
  \hline
\end{tabular}
\caption{\label{tab:design}Derived design principles on how to provide AI-generated instructions for students to write peer reviews.}
\end{center}
\end{table}

\begin{figure*}[!h]
    \centering
    \includegraphics[width=\linewidth]{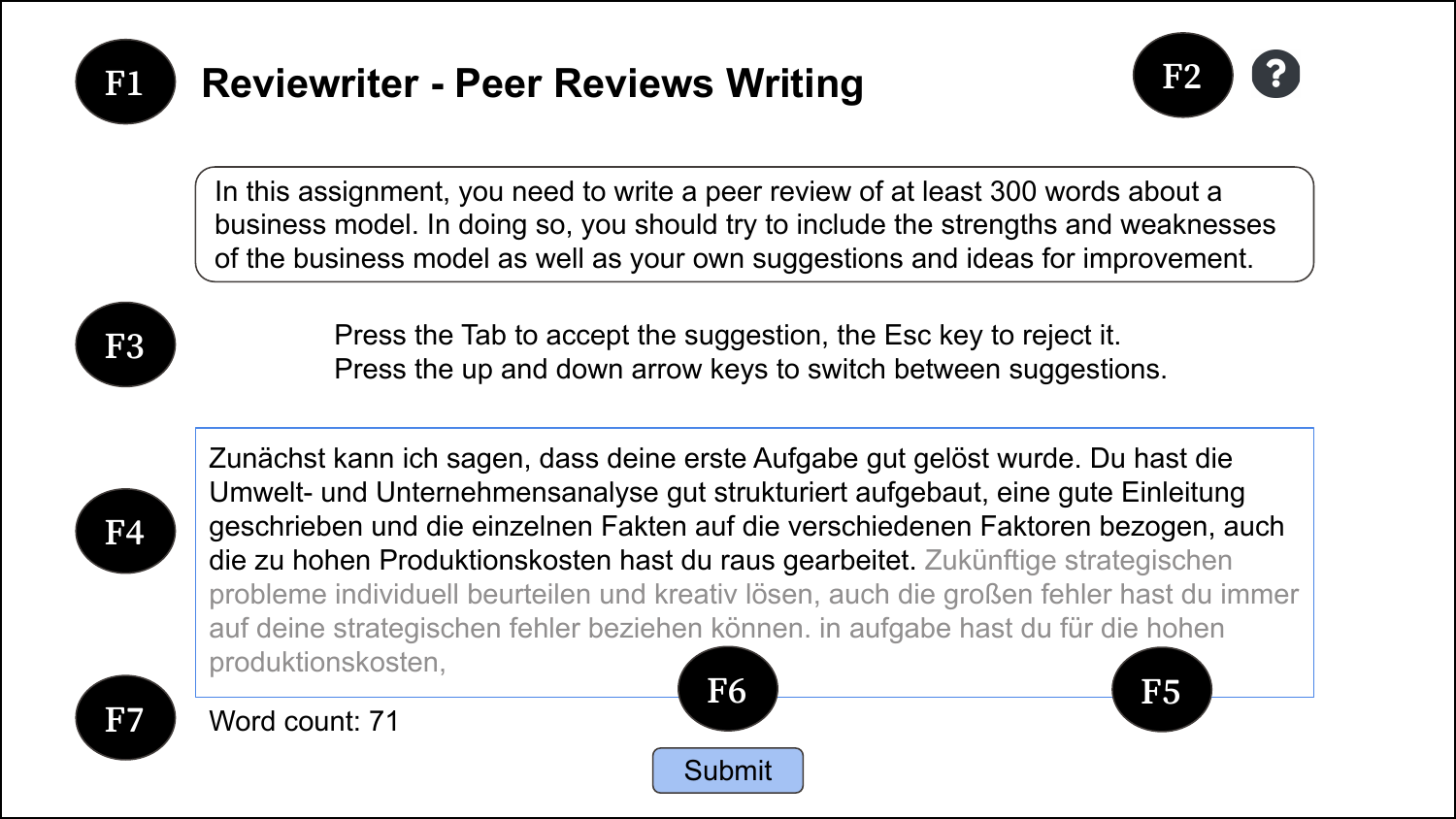}{
    \caption{A screenshot of \texttt{Reviewriter} and its main functionalities (\textit{F1} - \textit{F7}) derived from system requirements and design principles. The system provides a clean interface (\textit{F1}). By clicking the question mark, students get detailed guidance on the peer review writing task and the usage of the tool (\textit{F2}). A simple text area supports all typical interactions, such as typing, selecting, editing, and deleting text, and caret movement via keys and mouse (\textit{F4}). In the input area, the sentences in black are the actual text, we display the AI-generated instruction in an inline format in gray (\textit{F5}). The model generates next-sentence predictions to give students a complete view of the idea (\textit{F6}). We provide three instructions each time, and students may use the \textit{Tab} key to accept, the \textit{Esc} key to reject, and the \textit{Up} and \textit{Down} arrow keys to toggle through different instructions (\textit{F3}). The total number of words is displayed below the text area to inform students of their writing progress (\textit{F7}).}
    \label{fig:ui}
    }
\end{figure*}
With insights derived from the literature review and requirements from student interviews (similar to \citet{Rietsche.2018}), we develop seven design principles (Table \ref{tab:design}) and further map them to seven functionalities (Figure \ref{fig:ui} \textit{F1 - F7}) in \texttt{Reviewriter}, a responsive web application to provide AI-generated instructions for peer review writing. The design is student-centered and has two main components: a neat interface with key commands for text editing (Figure \ref{fig:ui}) and a generative language model in the backend \ref{sec:model}. To foster the independent thinking of students and discourage over-reliance on technology \cite{adiguzel2023revolutionizing}, we organize a workshop with two senior researchers to deliberate on the optimal timing for presenting the generated instructions. Combined with studies \citet{21multiple, bhat21people}, we decide to present instructions until students have entered a minimum number of words and put a certain amount of delay before showing instructions to minimize potential disruptions caused by irrelevant information from model hallucination \cite{maynez20faithfulness}. Figure \ref{fig:architecture} presents the system architecture. The student starts with writing the beginning of the review. The system will display instructions until students enter at least 25 words. After this threshold, when the student gets stalled, by pressing the spacebar, they will trigger the model in the backend to generate instructions. After the keypress, there is a delay of eight seconds before they receive instructions. To preserve the context while avoiding too much overhead for querying the mode, we pass the last twenty words from the input to the model. According to \textit{UR 4}, and supported by \citet{2020novelists}, overly brief suggestions are often unhelpful. To ensure clarity and concision, we limit each instruction to a maximum of 60 tokens, which is approximately 45 words \footnote{https://help.openai.com/en/articles/4936856-what-are-tokens-and-how-to-count-them}. In their experiment with one, three, and six instructions, \citet{21multiple} discovered that multiple instructions can facilitate the identification of useful phrases and boost their acceptance rate. We decide to present three instructions each time considering the cost-benefit tradeoffs for efficiency (e.g. reading time vs diversified content). The student controls the final output by checking multiple instructions and deciding whether to accept or reject them. They are free to add, delete, and replace the generated content.

\begin{figure*}[!h]
    \centering
    \includegraphics[width=\linewidth]{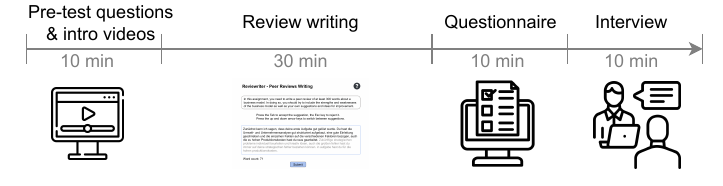}{
        \caption{Overview of the study procedure. Students begin with five pre-test questions and two introduction videos. Then, they engage in a 30-minute review writing task. Afterward, they are asked to complete a questionnaire, which is followed by an interview with a set of open-ended questions.}
        \label{fig:eval}
    }
\end{figure*}

\section{Evaluation of Reviewriter}
\subsection{Experimental setup}

To assess our prototype, we conduct a mixed-method study with fourteen students from a public university in Europe. We reach out to students who have participated in our previous design interview and also recruited students on campus. Fourteen students--eleven males and three females--participated in the evaluation. Three of them were undergraduate students and the rest were graduate students. Four graduate students also participated in our previous design interview. They were all native German speakers and expressed interest in getting AI-generated instructions when writing peer reviews. They have diverse backgrounds, including computer science, robotics, and business with a mean age of 25.33 years (SD = 3.60). The evaluation is conducted either face-to-face or remotely with a conference tool. Each student screen records their writing process, the interviews are also recorded and transcribed by a researcher.

\begin{enumerate}
    \item Pre-test (10 minutes): The experiment starts with a pre-survey that has five questions (Appendix \ref{appendix:pre}) followed by two videos. The first four questions measure the learners' level of innovation in the field of information technology, following \citet{agarwal2000time}. They need to rate their agreement with a statement on a Likert scale ranging from 1 (totally disagree) to 7 (totally agree), with 4 being neutral \cite{likert1932technique}. Following the pre-survey, we present two videos. The first video introduces a business model for a platform that connects ski instructors with learners, and the second video provides guidance on how to use \texttt{Reviewriter}.
    
    \item Peer review writing (30 minutes): In this phase, students are asked to write a review for a peer's business model. Specifically, they are asked to elaborate on strengths, weaknesses, and suggestions for improvement of the given business model. We instruct students not to use search engines and spend a minimum of 15 minutes on the task. A countdown indicates the remaining time.
    
    \item Questionnaire and interview (10+10 minutes): In the post-survey, we ask 29 questions (Appendix \ref{appendix:post}) to measure \textit{perceived ease of use}, \textit{perceived ease of interaction}, \textit{perceived level of enjoyment}, \textit{perceived level of excitement} and \textit{perceived usefulness}, following the technology acceptance model of \citet{venkatesh2008technology} and \citet{venkatesh2003user}. All constructs are measured with a 1- to 7-point Likert scale. Moreover, we ask several qualitative questions to further examine students' attitudes toward AI-generated instructions and capture the demographics.
\end{enumerate}

\subsection{Quantitative analysis and qualitative feedback}
\begin{table*}[!h]
\begin{center}
\begin{tabular}{ | c | p{0.12\linewidth} | p{0.15\linewidth} | p{0.15\linewidth} | p{0.15\linewidth} | p{0.12\linewidth} |}
\hline
Statistics & \textbf{Perceived ease of use} & \textbf{Perceived ease of interaction} & \textbf{Perceived level of excitement} & \textbf{Perceived level of enjoyment} & \textbf{Perceived usefulness} \\
\hline
\textbf{Mean} & 6.07 & 5.50 & 5.64 & 5.43 & 4.64\\
\hline
\textbf{Std.} & 0.83 & 1.22 & 1.15 & 1.16 & 1.34\\
\hline
\textbf{\makecell{Normalized \\mean}} & 0.87 & 0.79 & 0.81 & 0.78 & 0.66\\
\hline
\end{tabular}

\caption{\label{tab:results} Descriptive statistics from quantitative measure in the evaluation of \texttt{Reviewriter} (N=14). The measure of technology acceptance on a 1 - 7 Likert Scale (1: low, 4: neutral, 7: high).}
\end{center}
\end{table*}
 
To measure student perceptions of AI-generated instructions for peer review writing, we calculate the following constructs on a 1- to 7-point Likert scale  (Table \ref{tab:results}): perceived ease of use ($M_1=6.07, SD_1=0.83$), perceived ease of interaction ($M_2=5.50, SD_2=1.22$), perceived level of excitement ($M_3=5.64, SD_3=1.15$), perceived level of enjoyment ($M_4=5.43, SD_4=1.16$), and perceived usefulness ($M_5=4.64, SD_5=1.34$). The results show that the participants rate positively using \texttt{Reviewriter} to receive adaptive instructions. Moreover, the mean values of the tool are also very promising when comparing the results to the average of the scale. All results are better than the neutral value of four. This fosters motivation and engagement to use the learning application. \citet{malik2021adoption} found that perceived ease of use ($M_1=6.07$) and usefulness ($M_5=4.64$) positively influence student adoption intentions and their attitudes toward AI-based applications. The positive levels of perceived ease of interaction ($M_2=5.50$), excitement ($M_3=5.64$), and enjoyment ($M_4=5.43$) suggest that the technology has been accepted favorably. This is especially important for learning tools to ensure students are perceiving the usage of the tool as enjoyable, useful, and easy to interact with \cite{marangunic2015technology}. These are promising results for using this tool to receive AI-generated instructions in a peer review setting. 

In addition to quantitative scores, we incorporate qualitative open-ended questions to further understand student attitudes toward writing with AI-generated text and how the instructions impact their writing process. We translate the responses from German and cluster the representative ones (Appendix \ref{appendix:feedback}). The general attitude towards \texttt{Reviewriter} was very positive. Five students stated concretely the benefits of \texttt{Reviewriter} on their writing process. Three students mentioned the system is simple and easy to interact with. On the adoption of the generated instructions, one student used them every time, two students stated that they did not find anything useful in the instructions. Another two students reported that they never used the complete instructions but they picked up ideas or keywords from them. Five of them used instructions three to five times, and the rest stated that they use the AI-generated instructions quite frequently and did not provide an exact number. Moreover, it is interesting to note that there are divergent opinions on the delay of the system. Three students complained about the waiting time was too long while two other students were in favor of the delay and stated that the waiting time left them room to think. Finally, students enjoyed the diverse content in AI-generated instructions while noticing there were ungrammatical sentences and irrelevant phrases from time to time.

\section{Discussion}
Peer review writing is an increasingly important educational task in large-scale or distance learning scenarios since it enables personalized feedback to be delivered at scale, thereby lessening the workload of instructors \cite{er2021collaborative} and boosting learners' motivation \cite{16motivation}. However, during writing peer reviews, students may experience obstacles such as writer's block \citet{1980rigidblock} where they struggle to generate the next line, the right phrase, or the sentence \citet{oliver1982helping}. LLMs can help to overcome this obstacle by producing adaptive instructions based on students' input, which ultimately aid in the seamless progression of thoughts \cite{2022sparks}. To do so, we develop a novel peer review writing tool called \texttt{Reviewriter}. It allows students to use AI-generated instructions as an inspiration and incorporate those ideas into their own work in a creative and original way, such as by adapting, mixing, or reinterpreting those instructions \cite{qadir2022engineering}.

Our study contributes at least three key aspects to the innovative use of NLP in education. First, we explore the personalization of AI-generated instructions in a specific pedagogical scenario - peer review writing \cite{pardos2023learning} by gathering insights from literature review and student interviews \cite{verleger2018pilot}. Second, in contrast to \citet{lee2022coauthor} which used GPT-3 without adaptation for collaborative writing, we fine-tune three German language models on a corpus selected based on certain criteria to provide specialized content with high quality. Afterward, we choose German-GPT2 based on quantitative measures and human evaluation. Third, as noted in \citet{kuhail2023interacting}, "lack of feedback" is one of the challenges to using generative models in education. Therefore, we evaluate our tool with fourteen students and the result reveals positive technology acceptance based on quantitative measures. Through our qualitative evaluation, we find that students generally enjoyed seeing generated instructions with varied content to spark ideas. And they were enthusiastic and excited about writing with generative language models. We recognize that there is a need for further research on the effectiveness of LLM-based writing support tools in various contexts, as well as the improvement of faithfulness and factuality in AI-generated instructions \cite{maynez20faithfulness}. Nonetheless, our study contributes to the growing body of knowledge on the potential of generative AI to provide personalized writing instructions and enhance students' learning experiences \cite{pardos2023learning}.
\section{Conclusion and future work}
To help students mitigate writer's block during peer review writing, we design, build, and evaluate \texttt{Reviewriter}, a novel tool that aims to provide students with AI-generated instructions during their peer review writing process. We provide design insights with pedagogical considerations of integrating LLMs into peer review writing systems. Our evaluation involves fourteen students from tertiary education, who reported enjoying the interaction with the system, finding it easy to use, and expressing interest in using similar tools in the future. They also pointed out that the relevance of the generated instructions could be further improved. We present \texttt{Reviewriter}, including its design rationales and evaluated interface, as a contribution to the exploration of LLMs' potential in innovative NLP-based approaches in education. As NLP continues to advance, we aspire that our work will encourage other researchers to explore how generative AI can be integrated into educational applications to benefit teachers and students, while promoting responsible and ethical use.

For future work, we will investigate students' perceptions of peer reviews from different sources: their peers, peers using \texttt{Reviewriter}, and entirely AI-generated reviews. We will collect ratings and feedback from students who receive these reviews and compare the relevance, quality, and usefulness of the texts generated from each source. Additionally, we aim to integrate \texttt{Reviewriter} into the university's existing peer review system, enabling widespread adoption among students across various courses. By incorporating AI-generated instructions into routine peer reviews, we can examine the long-term impact on students' writing skills, critical thinking abilities, and overall academic performance. To enhance the relevance of the AI-generated instructions in \texttt{Reviewriter}, we will refine the algorithms and models based on feedback from our evaluation participants. Our iterative development process will involve incorporating more contextual information, employing advanced NLP techniques, and leveraging user feedback to achieve higher accuracy and helpfulness in the AI-generated instructions.

\bibliographystyle{ACM-Reference-Format}
\bibliography{ref}

\newpage
\appendix
\section{Details on data pre-processing and models}
\subsection{Template questions asked students to write reviews which some students copied to their review text}
\label{appendix:tq}
\begin{itemize}
    \item What do you see as the strengths of the fellow student's solution?
    \item What do you see as weaknesses in the fellow student's solution and how can they be addressed?
    \item What should be paid attention to in the revision of the solution?
    \item Provide concrete suggestions for improvement in this regard.
    \item Give concrete suggestions for improvement (constructive feedback).
    \item What should you pay attention to in the revision of the solution? Give concrete suggestions for improvement (constructive feedback).
\end{itemize}

\subsection{Abbreviations and expansions}
\label{appendix:abb}
\begin{table}[h]
\centering
\begin{tabular}{|c|c|}
    \hline
    Abbreviation & Expansion \\
    \hline
    bsp, bspw & beispielsweise \\
    \hline
    dh & da her \\
    \hline
    ev, evtl & eventuell \\
    \hline
    ggf & gegebenenfalls \\
    \hline
    oä & oder ähnliches \\
    \hline
    vlt & vielleicht \\
    \hline
    zb & zum Beispiel \\
    \hline
\end{tabular}
\caption{A list of abbreviations students used in the review text and we replace with the expansion in the pre-processing.}
\label{tab:feat}
\end{table}

\subsection{Hyperparameters for pretrained language models}
\label{appendix:hyper}
\begin{table}[!h]
\begin{center}
\begin{tabular}{|c|c|c|}
\hline
Hyperparameter & GPT2 & BLOOM \\
\hline
Vocabulary size & 50257 & 250880 \\
\hline
Attention heads & 12 & 8 \\
\hline
Hidden layers & 12 & 2 \\
\hline
Attention dropout & 0.1 & 0.1 \\
\hline
\end{tabular}
\caption{Hyperparameters for pretrained GPT2 and BLOOM}
\end{center}
\end{table}

\subsection{Sample text generated by different language models}
\label{appendix:text_en}
\begin{table}[!h]
\begin{center}
\begin{tabular}{ | p{0.2\linewidth} | p{0.7\linewidth} | } 
  \hline
  Source & Text \\ 
  \hline
  Original input & First of all, I can say that your first task was solved well. You have structured the environmental and company analysis well, written a good introduction, and related the individual facts to the various factors, and you have also worked out the excessively high production costs. \\ 
  \hline
  German GPT-2 generated instructions & Furthermore, the problems were identified and their elaboration was well coordinated. However, the behaviour of this company is completely unnecessary and one could hardly judge it more. Unfortunately, you went into far too much detail about the individual points and did little about the environmental and company analysis. 2) What do you see as the weaknesses of the environmental and business analysis and how can these be remedied?1.written out above:However, with the exception of case-related, these are other generic terms that have a rather different linguistic form. \\
  \hline
  BLOOM generated instructions & I have already recognized the second task and the reasoning is also understandable for you. Good luck for the future! Strengths) The project idea is cleanly and realistically implemented, as it is a \\ 
  \hline
  GerPT2-large generated instructions & You mastered the task very well. I have a few suggestions for improvement. First of all, you could have explained in the assignment what an environmental and company analysis actually is. I would also replace the word "company" with "firm" or something similar. You answered task two well, but you might have briefly explained the stakeholder approach, which was necessary to understand the task. Perhaps you could have changed the focus of the company analysis, but that is only a small point. In general you could \\
  \hline
\end{tabular}
\caption{\label{tab:text}Comparison of the sample text generated by different transformer-based pretrained language models with \texttt{max\_length=150} (we translated all text from German to English for the purpose of this paper).}
\end{center}
\end{table}

\section{Details on evaluations}
\subsection{Pre-test questions asked during evaluation of \texttt{Reviewriter}}
\label{appendix:pre}
\begin{enumerate}
    \item I like experimenting and trying out new technologies.
    \item As a rule, I am hesitant when trying out new technologies.
    \item In my circle of friends, I'm usually the first person to try new digital media / new technologies.
    \item When I hear about new technologies I look for a way to experiment with them.
    \item I have had experience writing reviews/feedback in the past.
\end{enumerate}

\subsection{Post-test questions asked during evaluation of \texttt{Reviewriter}}
\label{appendix:post}
\begin{itemize}
    \item Transition questions: How many times have you accepted \texttt{Reviewriter}'s recommendations?
    \item Technology Acceptance Model
    \begin{enumerate}
        \item Assuming the review writing assistance tool is available, the next time I want to write a review/feedback I would use it again.
        \item With \texttt{Reviewriter} I can write reviews/feedback more effectively.
        \item Learning to use \texttt{Reviewriter} was easy for me.
        \item I find using \texttt{Reviewriter} useful for writing reviews/feedbacks.
        \item I find \texttt{Reviewriter} easy to interact with.
        \item It would be easy for me to become familiar with \texttt{Reviewriter}.
        \item Compared to other participants, I think I wrote a very convincing review/feedback.
        \item After using \texttt{Reviewriter}, my ability to write reviews/feedback has improved.
        \item I'm sure I wrote a very insightful review/feedback.
        \item I'm sure I wrote a very convincing review/feedback.
        \item With \texttt{Reviewriter} I can write better reviews/ feedbacks.
        \item I think I now know more about how to write well-structured, persuasive, and insightful reviews/feedbacks.
        \item Assuming \texttt{Reviewriter} was available, the next time I write a review/feedback I would use it.
        \item After using \texttt{Reviewriter}, my ability to pay attention to the different parts of the review/feedback structure has improved.
    \end{enumerate}
    \item Evaluate student perceptions on the AI-generated instructions
    \begin{enumerate}
        \item I expect \texttt{Reviewriter} will help me improve my ability to write well-structured reviews/feedbacks.
        \item I assume \texttt{Reviewriter} would help me improve my ability to write compelling reviews/feedback.
        \item I assume \texttt{Reviewriter} would help me improve my ability to write insightful reviews/feedback.
        \item Interacting with the tool was fun and enjoyable for me.
        \item I expect \texttt{Reviewriter} will help me improve my ability to write helpful reviews/feedback.
        \item Interacting with the tool was exciting.
    \end{enumerate}
    \item Open-ended questions for qualitative feedback
    \begin{enumerate}
        \item How has \texttt{Reviewriter} impacted your writing process?
        \item What did you particularly like about using \texttt{Reviewriter}?
        \item Do you have any other ideas?
        \item What could still be improved?
        \item Have you used a writing support program before (e.g. Grammarly)?
        \item What is your field of study?
        \item Please enter your gender.
        \item Please indicate your mother tongue.
    \end{enumerate}
\end{itemize}

\subsection{Clustered qualitative student feedback from the evaluation of \texttt{Reviewriter}}
\label{appendix:feedback}
\begin{table*}[!h]
\begin{center}
\begin{tabular}{ |c|c|c|} 
  \hline
  Topic & Cluster & Statement \\ 
  \hline
  \multirow{2}{*}{\makecell{On the adoption of \\ the AI-generated \\ instructions}}
    & Positive & \makecell{S1: "I mainly accepted the ideas and slightly rewrote the \\proposed text." \\
    S3: "I find myself be inspired by professional keywords." \\
    S11: "I used the recommendations every time."}
    \\\cline{2-3}
    & Constructive & S4: "Never. They were utterly useless." \\
  \hline
  \multirow{2}{*}{\makecell{On the quality of \\ the AI-generated \\ instructions}}
    & Positive & \makecell{
    S1: "A few of the suggested ideas were very relevant. \\It also often remind me to say something positive." \\
    S4: " I like that it suggests diverse ideas that are quite \\different from each other." \\
    S10: "Reviewriter provided me with novel ideas that I could \\explore in depth."
    }
    \\\cline{2-3}
    & Constructive & \makecell{
    S1: "Shorter instructions would be more relevant sometimes." \\
    S10: "The instructions sometimes have spelling mistakes." \\
    S11: " Sometimes I got instructions that didn't fit the content." \\
    S12: "I would suggest to generate shorter snippets. \\Sometimes the beginning wasn't bad but later it got weird."} \\
  \hline
  \multirow{2}{*}{\makecell{On the impact of \\ the writing process}}
    & Positive & \makecell{
    S2: "The tool helps break through writer's block." \\
    S3: " When I got stuck on what to write, it sometimes had \\useful keywords, which made me a little quicker." \\
    S10: "The review writing process has accelerated." \\
    S11: "I got new ideas from Reviewirter's suggestions. \\ I think the system not only helps to write structured reviews, \\ but also to come up with new ideas. \\ This is where I see the greatest potential." \\
    S14: " I didn't feel so alone while writing." \\
    }
    \\\cline{2-3}
    & Constructive & \makecell{
    S1: "Waiting for suggestions slowed down my writing process." \\
    S12: "I tried to adopt the instructions a couple of times \\to be more efficient. However, since the waiting time for the \\ instructions is very long, the process has been delayed."} \\
  \hline
  \multirow{2}{*}{\makecell{On the system \\ interaction}}
    & Positive & \makecell{
    S5, S8: "It is easy to use and simple to operate." \\
    S10: "It is easy to use and saves time." \\
    S11: "I liked that I was not forced to accept the instructions \\and I could choose among several options."}
    \\\cline{2-3}
    & Constructive & \makecell{
    S11: "I think it would be better if we could select the \\instructions with the mouse."
    } \\
  \hline
  \multirow{2}{*}{\makecell{On the delay \\ of instructions}}
    & Positive & \makecell{
    S2: "Latency is moderate." \\
    S9: "I did not get suggestions instantaneously, I really just \\got it when I wanted it. That was really good, \\because that way my thoughts did not get interrupted." \\
    S14: "It is good that the instructions don't come immediately \\after I stop writing. It didn't disrupt my flow of writing."}
    \\\cline{2-3}
    & Constructive & \makecell{
    S6: "The proposals come too late, \\I almost come up with my own ideas." \\
    S1, S10, S12: "The waiting time for suggestions is long."} \\
    \hline
\end{tabular}
\caption{We have categorized the qualitative feedback received from fourteen students (referred to as S1 to S14) from tertiary education, who participated in the evaluation of \texttt{Reviewriter}. We collected the feedback through open-ended questions in the post-survey and concluding interview. For qualitative questions answered in German, we translated the written responses into English. The interview was conducted in English, recorded with the students' consent.}
\end{center}
\end{table*}

\end{document}